\documentclass[12pt]{elsarticle}

\usepackage{graphicx}

\usepackage{graphics}
\usepackage{amsmath}
\usepackage{psfrag}
\usepackage{graphicx}
\usepackage{amsthm}

\theoremstyle{definition}

\newcommand{\dd}{\,\mathrm{d}}

\begin{document}

\begin{frontmatter}
	\title{Inefficiency of the Brazilian Stock Market: the IBOVESPA Future Contracts}
\author[Rocha Filho]{Tarc\'\i sio M.\ Rocha Filho}
	\ead{marciano@fis.unb.br}
\author[Rocha]{Paulo M.~M.~Rocha}

\cortext[cor1]{Corresponding author}

\address[Rocha Filho]{Instituto de F\'\i{}sica and International Center for Condensed Matter Physics,
	Universidade de Bras\'\i{}lia, 70919-970 Bras\'\i{}lia-DF, Brazil}
\address[Rocha]{Instituto de F\'\i{}sica, Universidade de Bras\'\i{}lia, 70919-970 Bras\'\i{}lia-DF, Brazil}

\begin{abstract}
	We present some indications of inefficiency of the Brazilian stock market based on the existence of strong
	long-time cross-correlations with foreign markets and indices. Our results show a strong dependence on foreign
	markets indices as the S\&P 500 and CAC 40, but not to the Shanghai SSE 180, indicating an intricate interdependence.
	We also show that the distribution of log-returns of the Brazilian BOVESPA index has a discrete fat tail in the time
	scale of a day, which is also a deviation of what is expected of an efficient equilibrated market.
	As a final argument of the inefficiency of the Brazilian stock market,
	we use a neural network approach to forecast the direction of movement of the value of the IBOVESPA future contracts,
	with an accuracy allowing financial returns over passive strategies.
\end{abstract}
\end{frontmatter}

\section{Introduction}

Weather stock markets are predictable goes back at least to 1900 when the French Physicist Louis Bachelier presented his
Doctoral thesis entitled {\it Th\'eorie de la sp\'e\-cu\-la\-tion}, where he developed a mathematical model describing market
prices movements as a random walk process, as exemplified by French government bonds~\cite{bachelier}.
More recent progress characterize such price motions as martingales or Wiener processes~\cite{merton,bouchaud}.
If so, then markets are inherently unpredictable and forecasting is impossible.
This is maybe the first enunciation, although not equivalent to, of what would become the Efficient Market Hypothesis (EMH),
formulated formally by Samuelson~\cite{samuelson}, who posit it as a formal Theorem stating that ``Properly Anticipated Prices
Fluctuate Randomly'', and Fama~\cite{fama1,fama2},
who verified its validity from empirical data modulo some reasonable assumptions.
It can be succinctly formulated in the statement that market prices always fully reflect all available information.
Lo and MacKinlay~\cite{lo1} rejected the random walk hypothesis from empirical data, and a lot of evidence is now
available showing that stock market prices have inherent long time correlations~\cite{lo2}. One point to be noted is the meaning
of ``available information'' or ``Properly Anticipated'', in the sense that not all agents have the same capacity of extracting information from markets,
and therefore the information available to a given agent is not necessarily the same available to another agent.
For instance, a technological innovation may offer insights on market data not previously available, and thus giving strategic advantage
to those using it. Additional discussions on the history and the long-standing debate on the EMH see~\cite{read,schleifer,malkiel,malkiel2,sewell}.

As a matter of fact, in a market in perfect equilibrium, where all information that can be extracted from the market
is reflected in the prices, no arbitrage profits are possible, and no market gains would be observed other than those earned from pure chance~\cite{buffet}.
This seems to be at odds with what market operators expect and experience, as pointed out by Grossaman and Stiglitz~\cite{grossman}.
The efficient market hypothesis must consequently be envisaged an idealized situation describing a specific market, with differing
accuracy for different places and different times. This accuracy can, in many instances, be very good, at least for some time span. 
but, nonetheless, not fully exact. The key point is that trade-offs between risks and expected returns are necessary for the workings
of any financial market~\cite{lo3}, and the presence of correlations, even weak, allow for some degree of forecasting.

The main goal of the present paper is to show that the Brazilian stock market, specifically the future
contracts of the BOVESPA index, can be predicted with enough accuracy to allow a significant profitability.
The underlying reasons are also discussed in the paper,
particularly why the EMF is only an approximation for the Brazilian stock market.
The literature on forecasting techniques in final markets is,
of course, huge, and it is beyond the scope of the present paper to review it in any significant way, so we refer the reader to the
References~\cite{box,shadbolt,atsalakis,wang,oliveira, chong} and references therein.
We adopted neural networks as forecasting method, ans this mainly based on the existence of the universal approximation theorem (see below),
that guarantees that if future values are a function of some judiciously chosen past values, then a neural network exists that approximates
this function with arbitrary precision. We must note that usually many
forecasting approaches described in the literature, and in particular
for neural networks, do not present explicitly the projected financial returns, which are given here. 

The structure of the paper is the following:
in Section~\ref{data} we present the data used in our analysis. Section~\ref{randsec} presents our results on
correlations in the Brazilian market, showing that they deviate from a usual random walk,
and also that the BOVESPA is long-range time cross-correlated to the US and French markets, but quite intriguingly not to Chinese market.
In Section~\ref{returnssec} we discuss the statistical distributions of log-returns
of the IBOVESPA future contract, evidencing a slight deviation from Gaussianity,
while in Section~\ref{forecastsec} we show that it is indeed possible to extract information from the past
with financial gain in the future. Section~\ref{concludsec} is devoted
to a summary and a brief discussion of our results, and some concluding remarks.

\section{Data Used}
\label{data}

We used the following publicly available data on internet Web sites used, for the period from
December, 28 2015 to September, 26 2018, all closing market values:
\begin{itemize}
        \item {\it Markets}: IBOVESPA Future (B3 - the Brazilian stock market), number of negotiated contracts of the IBOVESPA Future,
                BOVESPA index and BOVESPA volatility,
                S\&P 500, VIX - CBOE Volatility Index, NASDAQ, Dow Jones Industrial Average,
                Huaan Shanghai Composite, Nikei 225 and CAC 40.
        \item {\it Stocks}: Petrobras PN, Banco do Brasil ON, Ambev ON, Itaú-Unibanco PN , Bradesco PN and Vale ON.
        \item {\it Commodities}: Crude oil and Gold.
	\item {\it Currency}: Future contract for the exchange rate from American Dollar to Brazilian Real (Dolk19).
\end{itemize}

\section{Random walks and correlations in the Brazilian stock market}
\label{randsec}

\begin{figure}[ptb]
\begin{center}
	\scalebox{0.27}{\includegraphics{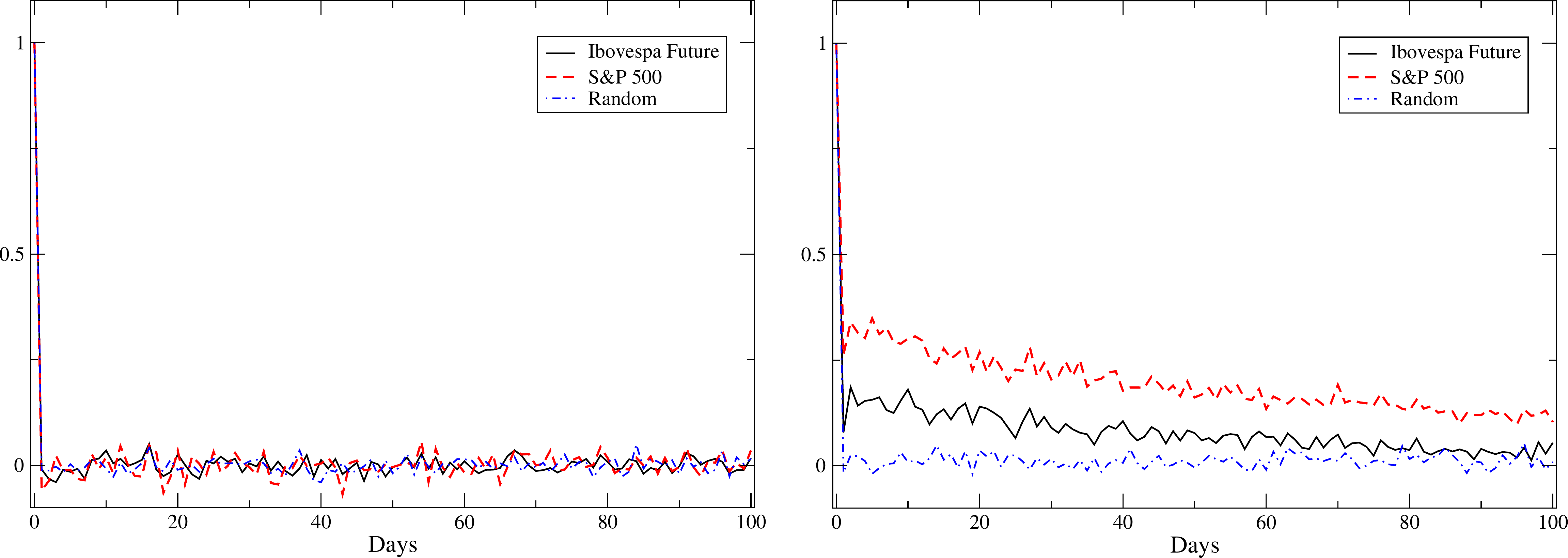}}
\end{center}
	\caption{Left: Auto-correlation function for the reduced log-returns as a function of the number of days for the IBOVESPA Future contracts,
	the S\&P 500 index and a random process.
	Right: Auto-correlation function for the absolute values of the reduced log-returns. Data are from January 4, 2000 to October 11, 2018.}
\label{fig_corrfuncs}
\end{figure}

\begin{figure}[ptb]
\begin{center}
        \scalebox{0.3}{\includegraphics{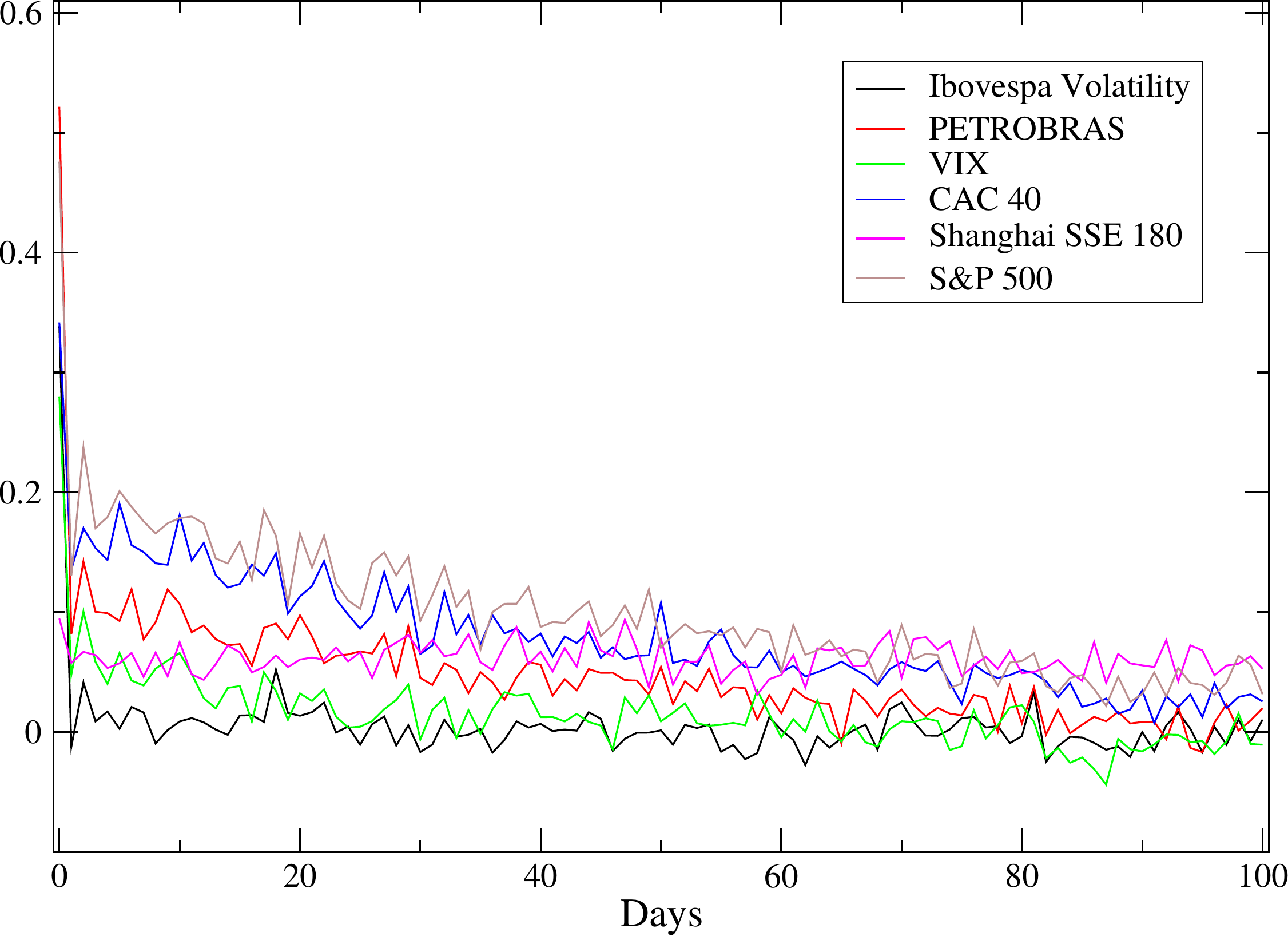}}
\end{center}
	\caption{Cross-correlations for the absolute value of the reduced log-returns of the
	IBOVESPA Futures contract with some stock prices and indices.}
\label{fig_others}
\end{figure}

We define the log-return $y_k(i)\equiv\log(x_k(i))-\log(x_k(i-1))$ with $x_k(i)$ the price of the asset $k$ at day $i$.
The reduced series $\tilde{y}_k$ are defined by:
\begin{equation}
	\tilde{y}_k(i)\equiv\frac{y_k(i)-\langle y_k\rangle}{\sigma_k},
	\label{redser}
\end{equation}
where $\langle\cdots\rangle$ stands for the average over the available data and $\sigma_k$ is the standard deviation of $y_k$.
The correlation functions are thence:
\begin{equation}
	C_{k,l}(n)=\left\langle\tilde y_k(i)\tilde y_l(i+n)\right\rangle.
	\label{corrdef}
\end{equation}
We also consider the correlations $\overline{C}_{k,l}(n)$ for the absolute values of the reduced log-returns $|\tilde y_k|$.
Figure~\ref{fig_corrfuncs} shows the auto-correlation functions for the IBOVESPA Futures, the S\&P 500 index, and a random walk
obtained from jumps generated from a uniform  uncorrelated random number generator,
with same variance as the IBOVESPA. For the log-returns there are no significant auto-correlation, while for the absolute
value of the log-returns long-range auto-correlations are evident, with the random walk auto-correlation at the noise level.
Figure~\ref{fig_others} shows the cross-correlation between the absolute values of the log-returns
of the IBOVESPA Futures and a few other indices and stock
values. Quite interestingly, no cross-correlation is observed with the volatility of the BOVESPA Index, while it is clearly non-zero for the VIX.
For the other series strong long standing cross-correlations are observed.

Since some bias can be present in the time series, as for non-stationary series,
we also perform a detrended fluctuation analysis, as developed for that purpose in Refs,~\cite{peng,podobnik}.
Let us consider the time series $y_k(i)$ and the corresponding reduced series
$\tilde y_k(i)$. The integrated time series are defined by:
\begin{equation}
	S_{k,l}=\sum_{i=1}^l\tilde y_k(i),
	\label{intseries}
\end{equation}
For a given integer $n$, these integrated time series are then divided into overlapping series of length $n+1$ with the elements
$\overline{S}_{k,l}^{(j)}=S_{k,l+j-1}$, $j=1,\ldots,N-n$, $l=1,\ldots,n+1$. A local linear trend $R_{k,l}$ for the $k$ sub-series
can be obtained from a least-squares fit inside the box $l=1,\ldots,n+1$.
The detrended correlation function for the $k$-th and $k^\prime$-the series is then given by:
\begin{equation}
	F^2_{k,k^\prime}(n)=\frac{1}{N-n}\sum_{j=1}^{N-n}f^2_{dca}(n,j),
	\label{detdef}
\end{equation}
where
\begin{equation}
	f^2_{k,k^\prime}(n,j)=\frac{1}{n-1}\sum_{l=1}^{n+1}(\overline{S}_{k,l}^{(j)}-\overline{R}_{k,l})
	(\overline{S}_{k^\prime,l}^{(j)}-\overline{R}_{k^\prime,l}).
	\label{detdef2}
\end{equation}
If long-term correlations are present in the original series, then $F^2_{k,k^\prime}(n)$ obeys a power law
of the form $N^{2\lambda}$ for sufficiently large values of $n$.
This type of analysis was used in Ref.~\cite{lima} to study correlation of the IBOVESPA index with its constituent stocks and related indices,
such as the negotiated volume and the change rate of the Brazilian Real to the US Dollar.

\begin{figure}[ptb]
\begin{center}
        \scalebox{0.3}{\includegraphics{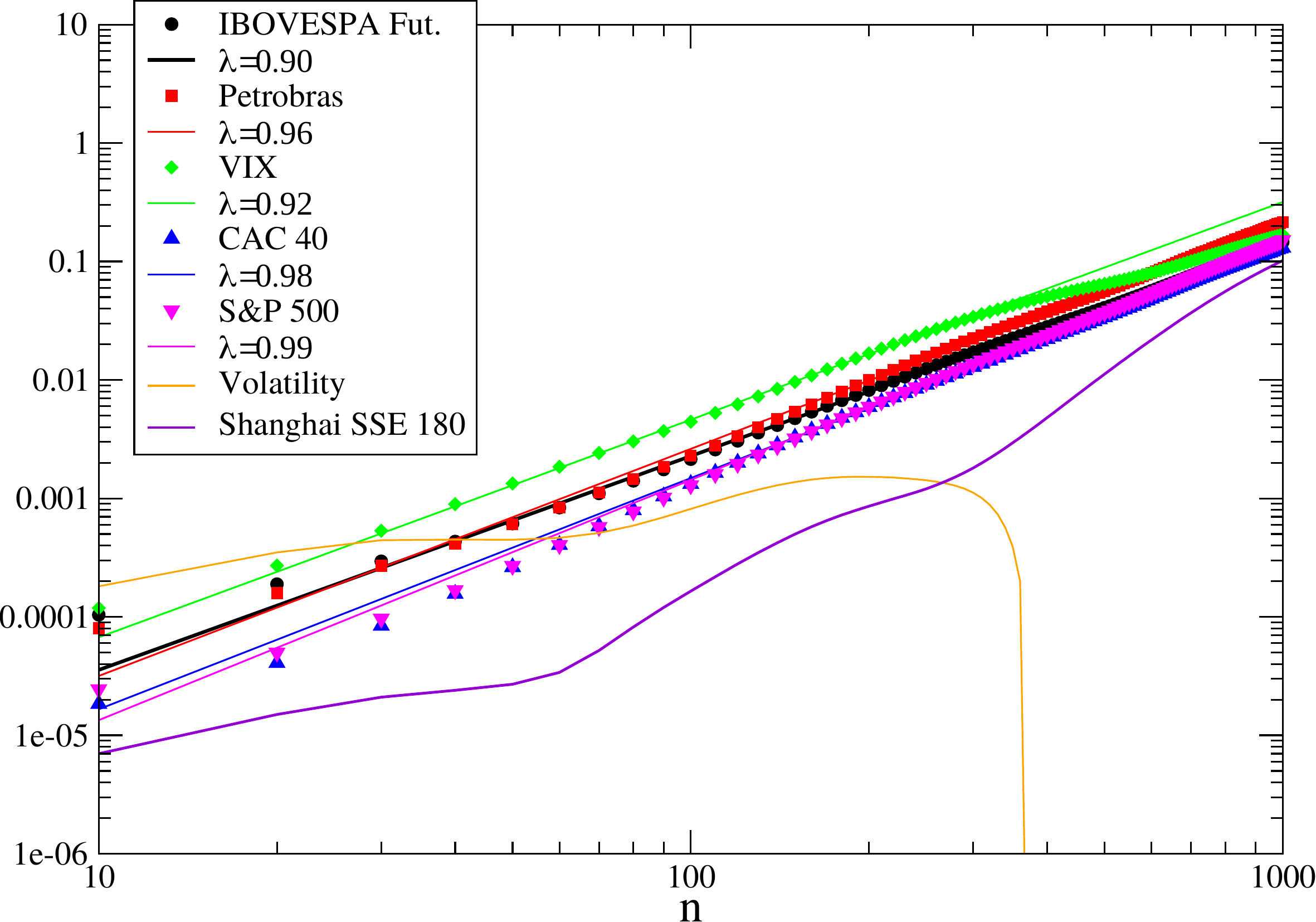}}
\end{center}
	\caption{Detrended correlations $F^2$ in Eq.~(\ref{detdef}) of the IBOVESPA Future contract with a few series.
	No power law was observed for the IBOVESPA volatility and the Shanghai Composite index.}
\label{fig_detrended}
\end{figure}

\begin{figure}[ptb]
\begin{center}
        \scalebox{0.28}{\includegraphics{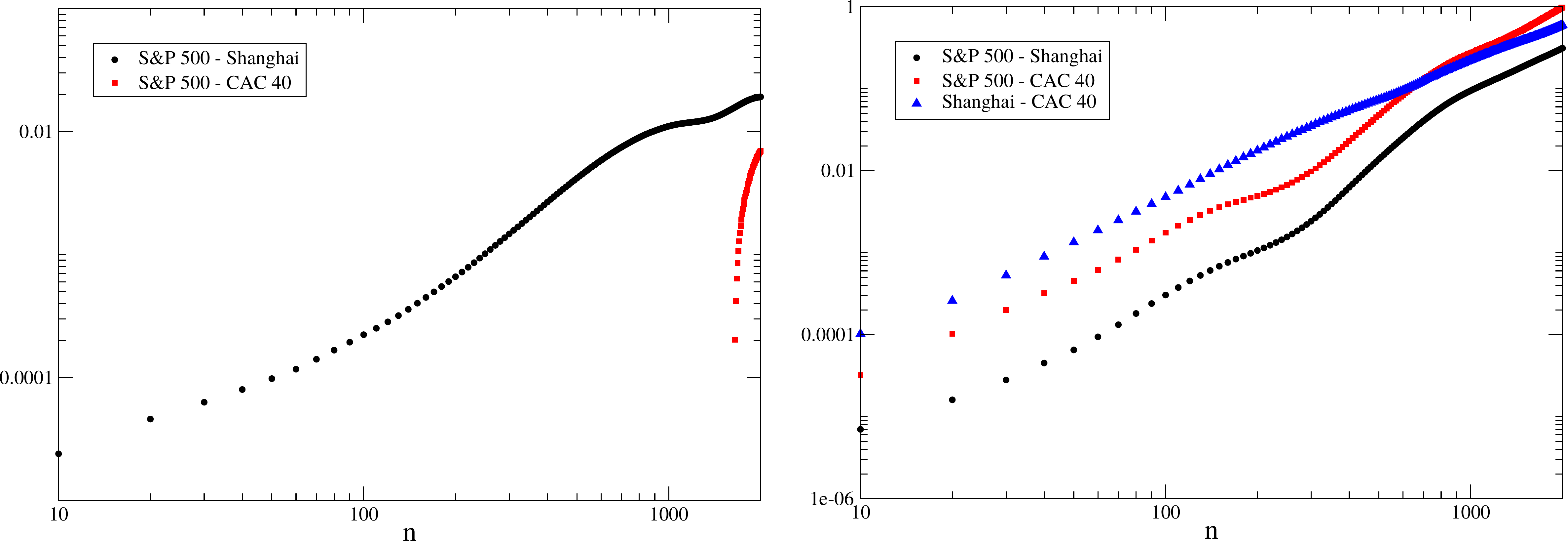}}
\end{center}
        \caption{Left panel: Detrended cross-correlation between the log-return of the S\&P 500 and Shanghai Composite.
	The values not shown for the S\&P 500-CAC 40 cross-correlation have negative values,
        as is the case for all values of the Shanghai Composite-CAC 40 cross-correlation.
        Right panel: same as the left panel but for the absolute values of the series.}
\label{fig_diffs_dcorr}
\end{figure}

Figure~\ref{fig_detrended} shows the detrended cross-correlations as defined in Eq.~(\ref{detdef}) for the IBOVESPA with some
stock and indices values. A power law is evidences for sufficiently large values of $n$. We observe a long-term cross-correlation
between the Brazilian and the US (S\&P 500) and French (CAC 40) markets, but not with the Chinese (SSE 180) market.
The IBOVESPA is cross-correlated to the VIX
volatility index but not to the volatility of the Brazilian (IBOVESPA) stock exchange itself. A more close look at these differences
is certainly worth of future research, and points to an intricate market interdependence. This is stressed by
Figure~\ref{fig_diffs_dcorr}, that shows the cross-correlation between the log-returns and the absolute values of the log-returns
between the S\&P 500, CAC 40 and SSE 180 indices. Clearly no long-term cross-correlation exists, contrary to
what is observed for the IBOVESPA-S\&P 500 and the IBOVESPA-CAC 40 cases. On  the other hand, the absolute
values of the log-returns are cross-correlated, as expected from the discussion above.

\section{Statistical distribution of returns}
\label{returnssec}

Another important feature of stock markets is the distribution function of the log-returns of commodities and indices.
It has been shown by Mantegna and Stanley~\cite{mantegna2} that high-frequency returns from the S\&P 500 index has fat tails,
and are well fitted by a truncated L\'evy distribution of the form~\cite{mantegna1}:
\begin{equation}
	f(x)=c\:L^{(\alpha)}(x),\:\:{\rm if} -d<x<d,\:\:{\rm and\:\:}f(x)=0\:\:{\rm otherwise},
	\label{truncdist}
\end{equation}
where $c$ is a normalization constant, $d$ is the value of the truncation and $L^{(\alpha)}(x)$ is the symmetric L\'evy
distribution with index $\alpha$, defined from its Fourier transform as:
\begin{equation}
	L^{(\alpha)}(x)=\frac{1}{\pi}\int_0^\infty\dd k\:{\rm e}^{-\gamma k^\alpha}\cos(kx),
	\label{levyfourier}
\end{equation}
where $\gamma>0$ is the scale length. The truncation is justified by the obvious fact that although high absolute values of returns
are indeed observed, no infinite returns are possible. Due to the central limit theorem (for a good historical review see~\cite{adams})
a random variable obtained from the sum of $n$ identical random variables with finite variance converges to a Gaussian in distribution,
as for the distribution $f(x)$ in~(\ref{levyfourier}). On the other hand if $l\rightarrow\infty$ then $f(x)$ becomes the
stable L\'evy distribution $L^{(\alpha)}(x)$. For $\alpha=1$ we obtain a Gaussian distribution. Although $f(x)$ converges asymptotically
to a Gaussian, its speed of convergence is usually very slowly (unless $d$ is small) and the distribution of the summed variables
stays close to a L\'evy distribution up to a crossover value for the number of summed variables,
that can be estimated, it approaches the Gaussian distribution according to the central limit theorem.
Another possible factor that can cause a slowdown of the convergence is the presence of correlations.
A procedure to determine the exponent $\alpha$ is described in Ref.~\cite{mantegna1}:
for a sum of $n$ identical and uncorrelated random variables $X_n=\sum_{i=1}^n x_i$,
where each $x_i$ has a distribution given in Eq.~(\ref{truncdist}),
the probability of return to origin $X=0$ is given for sufficiently small $n$ by:
\begin{equation}
	P(X_n=0)\approx\frac{\Gamma(1/\alpha)}{\alpha\pi n^{1/\alpha}},
	\label{retorigin}
\end{equation}
where $\Gamma$ is the Euler function. For large values of $N$, we obtain $P(X_n=0)\propto n^{-1/2}$.

\begin{figure}[ptb]
\begin{center}
        \scalebox{0.3}{\includegraphics{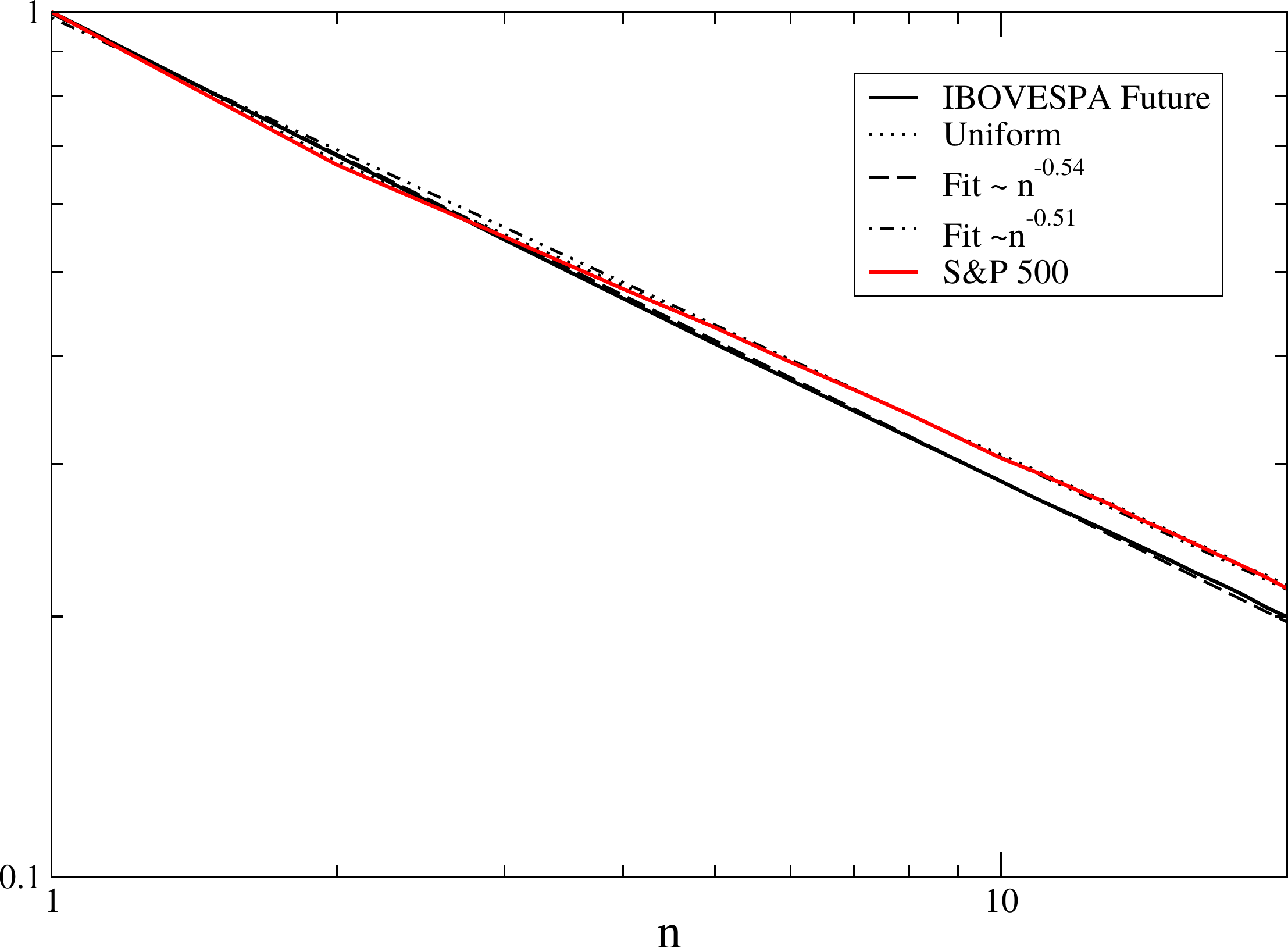}}
\end{center}
        \caption{Probability of return to zero for the log-returns for the BOVESPA index future contracts obtained from a random
        shuffling of log-returns from $4639$ opening days from January 4, 2000 to October 11, 2018 and for $5000$ realizations,
	for the S\&P 500 index and from the same number of points from a uniform distribution with zero mean and
	same standard deviation as the BOVESPA series.}
\label{fig_probzero}
\end{figure}

We use the procedure of the previous paragraph to test for fat tails in the
log-returns of the IBOVESPA Future contracts and the S\&P 500 stock index using the same data as in Fig.~\ref{fig_corrfuncs}.
The sum corresponding to $X_n$ is obtained by randomly shuffling the values in the series. This allows to eliminate any possible correlations
and to obtain many realizations of the underlying (and at least partially) stochastic processes. Figure~\ref{fig_probzero} shows the corresponding
results for both series with $5000$ realizations. For comparison purposes we also show the results for a random series with the same
standard deviation as the IBOVESPA Futures series. While the S\&P 500 index has a value of $\alpha$ compatible
with a Gaussian distribution, the IBOVESPA Futures has a value of $\alpha=0.54$, indicating the existence of a fat tail,
and this for inter-day data, implying that the Brazilian stock market is not yet equilibrated up to this time scale.

\section{Forecasting the IBOVESPA Future contract}
\label{forecastsec}

As discussed above, the Brazilian stock market has some peculiarities with respect to other more mature markets. Particularly,
a strong detrended cross-correlation with foreign markets as exemplified by the S\&P 500 and CAC 500.
On the other hand one can argue that despite any such deviations it is not possible to beat the market, i.~e.\ to devise a strategy
that yields higher returns than a passive investment.
If the efficient market hypothesis is valid for case at hand, no arbitrage is possible, i.~e.\ there is no way to beat the market
systematically. Conversely, if the hypothesis does not apply, then a reasonable amount of arbitrage should be possible.
We now show that it is possible to judiciously use available information to increase the profitability,
at least for the IBOVESPA future contracts, as we proceed to show.
For that purpose we use a feed-forward neural network as a forecasting tool using as input previous market indices and stock prices.

An important issue when looking for patterns in data
is to avoid spurious correlations that result from too large data sets~\cite{calude}.
This can spoil the forecasting capacity of the model, and thus we need to carefully
identify a data set not too large but yet conveying enough information which are expected to represent realistically, although approximately,
different variables that influence the target value to be predicted, and
is given by the series presented in Sec.~\ref{data}. Some choices may seem arbitrary, but are a result of a few simple logical
assumptions and some experimentation, and is certainly subject to possible further improvements.

\subsection{Feed-Forward Neural Networks}

Feed-Forward Neural Networks (FFNN) are composed by a number of layers of neurons organized as~\cite{nelis}:
an input layer, a number of hidden layers and an output layer. Neurons on each
layer are connected only to the neurons of the next layer by synaptic connections, each having a synaptic weight
denoted by $w^{(i)}_{jk}$ connecting the $j$-th neuron of the $i$-the layer to the
$k$-th neuron on the $(i+1)$-th layer. The output of $j$-the neuron of the $i$-the layer is then be written as
\begin{equation}
	y^{(i)}_j=\phi^{(i)}(z_j^{(i)}),\hspace{5mm}z_i^{(i)}=\sum_k w^{(i-1)}_{kj} y^{(i-1)}_k,
	\label{eq1}
\end{equation}
where $z_j^{(i)}$ denotes the activation of the neuron and $F^{(i)}$ is the activation function of the layer $i$.
Different choices for the activation function are possible, and examples include the logistic function~\cite{silva}:
\begin{equation}
	\phi(z)=1/(1+\exp(-z)),
	\label{eq2}
\end{equation}
the hyperbolic-tangent function
\begin{equation}
	\phi(z)=a\tanh(bz),
        \label{eq3}
\end{equation}
where $a$ and $b$ are adjustable parameters, and the linear function:
\begin{equation}
        \phi(z)=z.
        \label{eq4}
\end{equation}
Besides, it is useful to have for each layer (except for the output layer) an extra bias neuron
taking no input and unit output.
The main advantage of the FFNN is that it is an universal approximator as any continuous function can be approximate with
arbitrary precision with a FFNN with single hidden layer~\cite{hornik}.

\subsection{Principal Component Analysis}

In order to have a relatively simple structure for the neural network which allows a more efficient training,
the number of inputs must be limited. Thus some sort of dimensionality reduction must be considered, and a good choice
with a sound statistical foundation is the principal component analysis.
By considering the component with the greatest eigen-value we obtain a single time series,
instead of the 20 series listed in Sec.~\ref{data}.
In order to preserve a causality relation between the input and output series it is
important to keep the target series, the IBOVESPA future contracts,
in the set of series to used to compute the principal component.

Two different time windows of 5 and 10 days prior to the day of prediction are used to obtain the principal component,
which is then used as input for the neural network.
The principal component is obtained from the return series $\Delta x^{(i)}_k\equiv x^{(i)}_{k+1}-x^{(i)}_k$,
where $i$ specifies the original series $k$ the day in the sequence. The covariant matrix $M$ is given by its components:
\begin{equation}
	M_{i,j}=\left\langle \frac{(x^{(i)}-\overline{x}^{(i)})(x^{(j)}-\overline{x}^{(j)})}{\sigma^{(i)}\sigma^{(j)}} \right\rangle,
	\label{covmat}
\end{equation}
where $\langle\cdots\rangle$ denotes the average over the considered period and $\overline{x}^{(i)}$ and $\sigma^{(i)}$ stand
for the average and standard deviation of the $i$-th series over the same period, respectively.
By determining the eigen-vectors of $M$ we have a linear orthogonal transformation that converts the
original series into a set of (linearly) uncorrelated series, the principal components.
The series corresponding to the largest eigenvalue then yields the contribution with strongest
correlations.

\begin{figure}[h]
\begin{center}
\scalebox{0.35}{\includegraphics{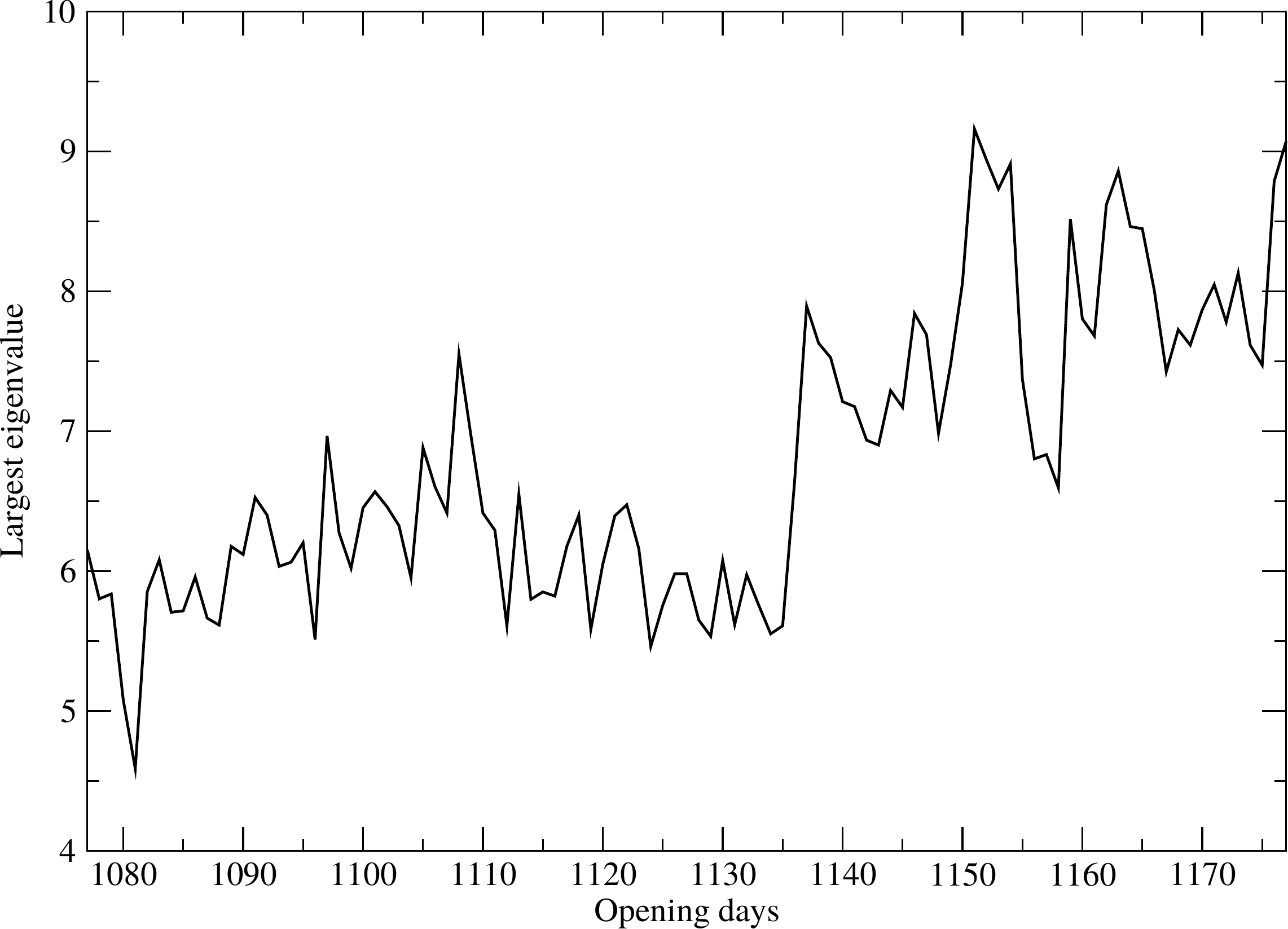}}
\end{center}
	\caption{Largest eigenvalue of the covariance matrix computed over the 10 previous days as a function of the day after the
	last day used in the computation, counting from December, 28 2015.}
\label{fig_largeigen}
\end{figure}

The time variation of the largest eigenvalue of the covariance matrix computed over a 10 days period
is shown in Fig.~\ref{fig_largeigen}, for the last one hundred days (final day used in the principal component)
of our time series. The components of the linear transformation resulting in the first principal component for the last day in the series
are given in Table~\ref{tab0}, and the set of all the 20 eigenvalues are given in Table~\ref{tab0b}.
Since the greatest eigenvalue is significantly greater than the other eigenvalues, except for the second and third,
the principal component retains much of the information and is thus a good representation of the set of data used,
resulting in a reduction of data dimensionality, from 20 to a single dimension. Although it possible to use more
than one component, this usually results in a more difficult training of the network and in a decrease of
its learning efficiency~\cite{chong}.

\begin{table}[h]
{\hspace{-20mm}\small
\begin{center}
\begin{tabular}{ c | c || c | c || c |c || c | c}
\hline
	Series N${}^o$ & Comp. & Series N${}^o$ & Comp. &Series N${}^o$ & Comp. & Series N${}^o$ & Comp. \\
\hline
	1 & -0.286 & 6  & -0.289 & 11 & -0.254 & 16 & -0.192\\
	2 & -0.291 & 7  & -0.285 & 12 &  0.198 & 17 & -0.248\\
	3 & -0.284 & 8  & -0.08  & 13 &  0.103 & 18 & -0.165\\
	4 & -0.28  & 9  & -0.235 & 14 & -0.071 & 19 &  0.162\\
	5 & -0.214 & 10 &  0.258 & 15 & -0.121 & 20 &  0.225\\
\hline
  \end{tabular}
\end{center}
	\caption{Components for each series for the first principal component for the last day in Figure~\ref{fig_largeigen}.}
\label{tab0}
}
\end{table}

\begin{table}[h]
{\hspace{-20mm}\small
\begin{center}
\begin{tabular}{ c | c | c | c }
\hline
	$-1.268\times 10^{-9}$ & $-9.053\times 10^{-10}$ & $-6.662\times 10^{-10}$ & $-5.343\times 10^{-10}$\\
	$-5.015\times 10^{-10}$ & $-4.111\times 10^{-10}$ & $-2.986\times 10^{-10}$ & $-5.153\times 10^{-11}$\\
	$2.454\times 10^{-12}$ & $4.008\times 10^{-10}$ & $4.323\times 10^{-10}$ & $6.591\times 10^{-10}$\\
	$7.82\times 10^{-10}$ & $9.553\times 10^{-10}$ & $1.23\times 10^{-9}$ & $0.52$\\
	$0.593$ & $3.091$ & $4.131$ & $11.665$\\
\hline
\hline
  \end{tabular}
\end{center}
	\caption{Eigenvalues of the covariant matrix for the last day in Fig.~\ref{fig_largeigen}.}
\label{tab0b}
}
\end{table}

\subsection{Prediction from the FFNN}
\label{predffnn}

To properly choose a FFNN with a good generalization property some experimentation is needed. Our best results were obtained
using 5 days of the time series as input, 30 neurons in a single hidden layer, and one output neuron for the predicted value for the IBOVESPA
future contract value one day ahead. Also, in order to limit the effect of noise, fluctuations and the effects of external causes,
we use a moving average over three days on the principal component series. The training is performed using a batch of 20 input-output data from
the days up to the day the prediction is performed (after market closure),
and by minimizing the square error of each output with respect to the accurate values for each element in the batch.
Different optimization algorithms were tested, such as the backpropagation, conjugate gradient and simplex algorithms~\cite{rochafilho,numrecip},
with the best choice being the ADAM stochastic optimization~\cite{adam}.

The decision on which position (hold or put) to assume using the result of the forecast is crucial for an efficient
methodology in a stock market. On the other hand, predicting the value of an asset, even within a given error,
is indeed very difficult due to the immense number of factor expected to influence. This is nonetheless not entirely the case
for the direction of the index movement (up or down). As explained above, we perform two different predictions using the data
from then principal component series, using five and ten days in each case.
If the resulting forecasts agree in the direction of the asset movement, a long or short position is taken accordingly.
Otherwise we take no position. We take as a significant indication of an upward movement if the output of the neural
network is above the threshold of $500$ points.

We also use the additional prescription of considering only long positions, which increases significantly the profitability
for the time period considered here.
This is justified by the usually obvious fact that the natural long term behavior of a stock market index is to increase in value,
which in our case amounts to say that a bullish market is easier to predict than a bearish market.
This will be justified pos hoc by our results. Since the market can change its mood over time a continuous follow up is required
in any practical implementation.

\begin{figure}[ptb]
\begin{center}
\scalebox{0.35}{\includegraphics{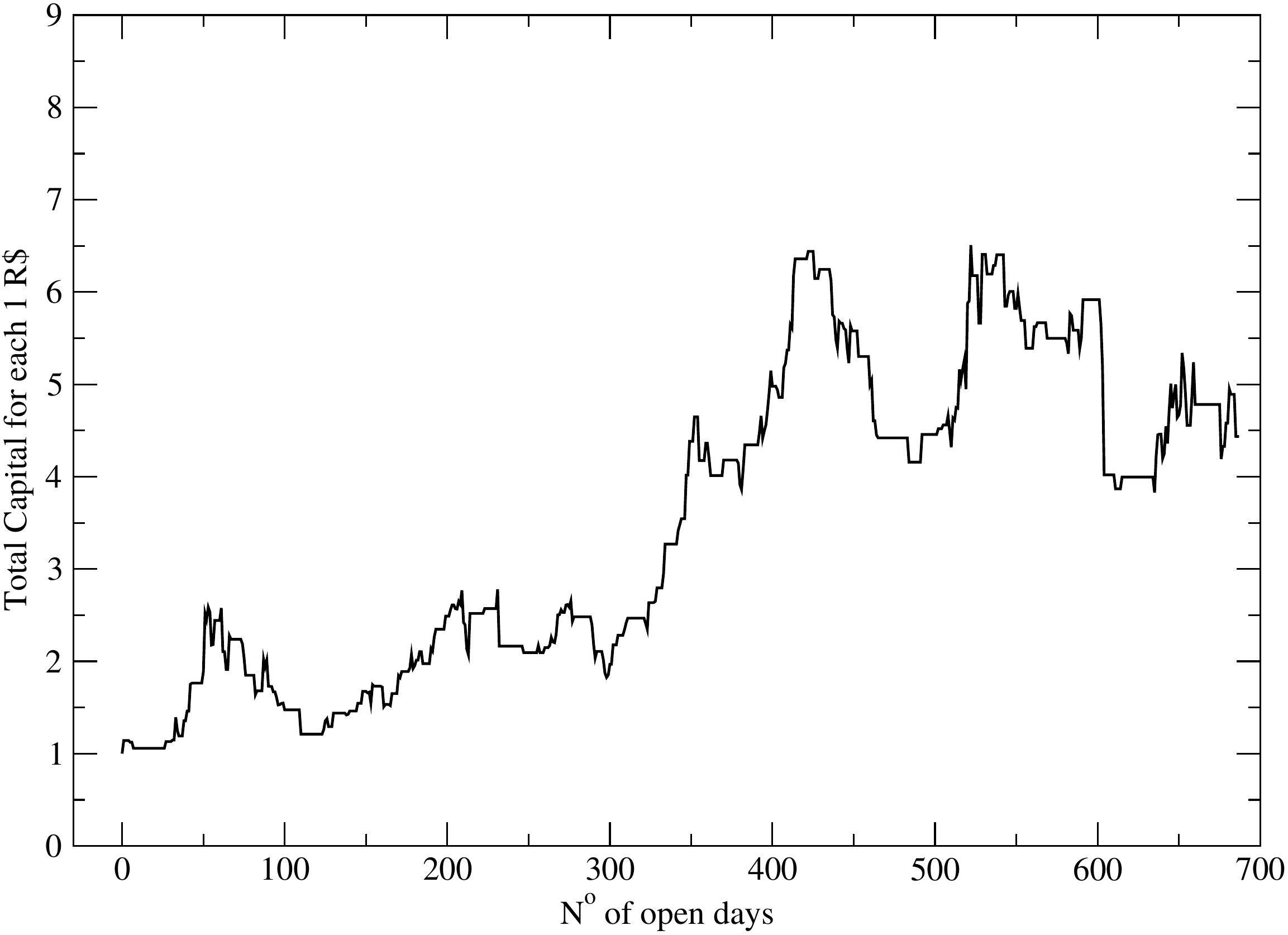}}
\end{center}
	\caption{Total capital for each $1$ R\$ without loss regularization with an average monthly return of $4.4\%$.}
\label{fig_gr_lucro}
\end{figure}

\begin{figure}[h]
\begin{center}
\scalebox{0.25}{\includegraphics{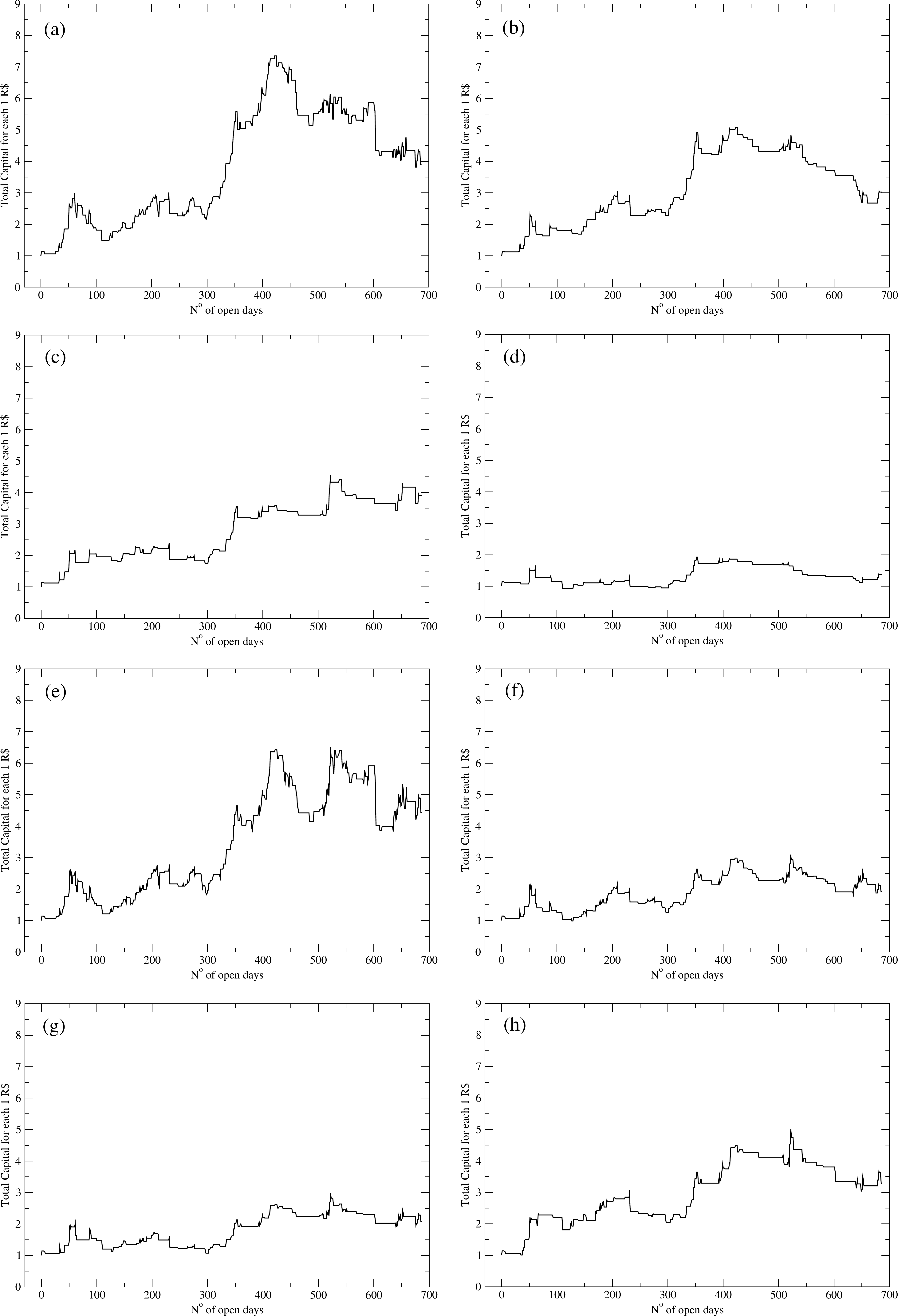}}
\end{center}
	\caption{Total capital with loss regularization. a) $N_l=1$, $H_g=0$; b) $N_l=1$, $H_g=1$; c) $N_l=1$, $H_g=2$; d) $N_l=1$, $H_g=3$;
	e) $N_l=2$, $H_g=0$; f) $N_l=2$, $H_g=1$; g) $N_l=2$, $H_g=2$; h) $N_l=2$, $H_g=3$.}
\label{fig_gr_lucro2}
\end{figure}

The time period used for the forecasting comprises 700 opening days of the B3 (BOVESPA) stock market in Brasil,
starting in December, 28 2015 and ending on October, 5 2018. The theoretical capital obtained from the present
strategy is shown in Fig.~\ref{fig_gr_lucro} for each unit in the original capital.
Although a high profit of $329.4\%$, the volatility is quite high. This can be mitigated, with a loss
in total return, by improving the decision strategy as follows. After $N_l$ days of loss all positions are zeroed and the
operations return after $N_g$ days of gain if investment using the algorithm had continued. Results of the capital evolution for different values of
$N_l$ and $N_g$ are shown in Fig.~\ref{fig_gr_lucro2}. To asses the volatility and other relevant information
we compute the percentile average day return $\overline{\Delta x}$ and the standard deviation $\sigma_{\Delta x}$ (volatility),
skewnes ${\cal{S}}_{\Delta x}$ and kurtosis ${\cal{K}}_{\Delta x}$ for the returns.
A combination of such strategies can be used along the time by examining and optimizing over previous results.

\begin{table}[h]
{\hspace{-20mm}\small
\begin{center}
\begin{tabular}{ c | c || c | c || c | c || c | c}
\hline
	$N_l$ & $N_g$ & Monthly & Total & $N_l$ & $N_g$ & Monthly & Total \\
\hline
	1 & 0 & 4.0\% & 290.1\% & 2 & 0 & 4.4\% & 343.6\% \\
	1 & 1 & 3.2\% & 202.4\% & 2 & 1 & 1.7\% & 92.6\% \\
	1 & 2 & 4.0\% & 290.3\% & 2 & 2 & 1.9\% & 106.8\% \\
	1 & 3 & 0.6\% & 36.8\% & 2 & 3 & 3.5\% & 228.1\% \\
\hline
  \end{tabular}
\end{center}
	\caption{Monthly average and total returns for a few values of $N_l$ and $N_g$.}
\label{tab1}
}
\end{table}

\begin{table}[h]
{\hspace{-20mm}\small
\begin{center}
\begin{tabular}{ c |c || c | c | c | c || c |c || c | c | c | c}
\hline
	$N_l$ & $N_g$ & $\overline{\Delta x}$ & $\sigma_{\Delta x}$ & ${\cal{S}}_{\Delta x}$ & ${\cal{K}}_{\Delta x}$ &
	$N_l$ & $N_g$ & $\overline{\Delta x}$ & $\sigma_{\Delta x}$ & ${\cal{S}}_{\Delta x}$ & ${\cal{K}}_{\Delta x}$ \\
\hline
	 1 & 0 & 0.0028 & 0.041 & 1.08 & 15.05 & 2 & 0 & 0.0019 & 0.044 & 0.9 & 11.11\\ 
	 1 & 1 & 0.0021 & 0.031 & 2.42 & 36.31 & 2 & 1 & 0.0014 & 0.037 & 1.35 & 18.54\\
	 1 & 2 & 0.0024 & 0.03  & 3.02 & 38.32 & 2 & 2 & 0.0016 & 0.034 & 1.68 & 24.34\\
	 1 & 3 & 0.0007 & 0.026 & 1.77 & 56.5 & 2 & 3 & 0.0022 & 0.031 & 2.56 & 35.6\\
\hline
\hline
  \end{tabular}
\end{center}
	\caption{Average $\overline{\Delta x}$, standard deviation $\sigma_{\Delta x}$, skewness ${\cal{S}}_{\Delta x}$ and
	kurtosis ${\cal{K}}_{\Delta x}$ of daily return $\Delta x$ for the same values of $N_l$ and $N_g$ as in Table~\ref{tab1}.
	The corresponding values for the straightforward FFNN approach in Fig.~\ref{fig_gr_lucro} are:
	0.0019, 0.049, 0.9 and 11.13, respectively.}
\label{tab2}
}
\end{table}

\section{Discussion and Conclusions}
\label{concludsec}

We presented indications that the Brazilian stock market has some degree of inefficiency, including a strong cross-correlation
with the US and French stock markets, and a small deviation from Gaussianity on the one-day time scale, not observed for
the US market. This suggests the possibility of extracting useful information from the financial series studied.
We implemented a Feed-Forward Neural Network to predict, one day ahead, the value of the IBOVESPA future contract, and considered
the direction of the movement as explained in Section~\ref{predffnn}. The choice is
guided by simplicity and the the universal approximation theorem for FFNNs, alongside the assumption that
a judicious choice of time series can provide a set of information on which a future stock value depends. If the value
of a stock or contract depends, even approximately and up to non-predictable external factors,
on (recent) past values of a few financial series, then a FFNN with a single hidden layer exists fitting with arbitrary accuracy
such a function. The precise structure of the neural network has to be determined on a trial-error basis. We discussed a procedure
for choosing this structure and how to design a decision strategy to partially control risk (volatility) and profitability
for the IBOVESPA future contracts. The predicted net returns are well over market yields from a passive strategy.
Of course much improvement is still in need for the present approach to be considered and attractive investment strategy,
and many directions are worth exploring, such as a more detailed analysis of the financial data used in the learning and
other dimensionality reduction procedures.
It is also relevant to observe that many drawbacks must be considered when performing this kind of analysis, as
spurious correlations and spurious patterns in data~\cite{smith}.
We believe that this is a direction for future research in forecasting the Brazilian and other emergent
financial markets. As a remark, we note that the detrended cross-correlated
analysis can be a powerful tool for assessing inter-market influence.

\section{Acknowledgments}

TMRF was partially funded by CNPq (a Brazilian government agency) under the Grant n0.\ 305842/2017-0.


\begin{thebibliography}{99}
	\bibitem{bachelier} Bachelier L.~(1900). {\it Th\'eorie de la Spéculation}.
		Annales Scientifiques de l'\`E.N.S.\ $3^e$ série 17, 21-86.
	\bibitem{merton} Merton C.~R.~(1973). {\it Theory of Rational Option Pricing}, The Bell Journal of Economics and Management Science,
		4(1), 141-183.
	\bibitem{bouchaud} Bouchaud J.~P.\ and Potters M.~(2003). {\it Theory of Financial Risk and Derivative Pricing: From Statistical
		Physics to Risk Management}. Cambridge Univ.\ Press.
	\bibitem{samuelson} Samuelson P.~A.~(1965). {\it Proof that Properly Anticipated Prices Fluctuate Randomly}.
		Management Review, 6(2), 41-48.
        \bibitem{fama1} Fama, E.~F.~(1965). {\it The behavior of stock-market prices}. The Journal of Business, 38(1), 34–105 .
	\bibitem{fama2} Fama E.~F.~(1970). {\it Efficient Capital Markets: A Review of Theory and Empirical Work}.
		The Journal of Finance, Vol. 25, No. 2, Papers and Proceedings of the Twenty-Eighth Annual Meeting
		of the American Finance Association New York, N.Y. December,28-30.
	\bibitem{lo1} Lo A.~W.~and MacKinlay A.~C.~(1988). {\it Stock Market Prices do not Follow Random Wals: Evidence
		from a Simple Specification Test}. Review of Financial Studies, 1(1), 41-65.
	\bibitem{lo2} Lo A.~W.~(1991). {\it Long-Term Memory in  Stock Market Prices}. Econometrica, 59(5), 1279-1314.
	\bibitem{read} Read C.~(2013). {\it The Efficient Market Hypothesists: Bachelier, Samuelson, Fama, Ross, Tobin, and Shiller}.
		Palgrave MacMillan.
        \bibitem{schleifer} Schleifer A.~(2000). {\it Inneficient Markets}. Oxford Univ.\ Press.
	\bibitem{malkiel} Malkiel B.~G.~(2003). {\it The Efficient Market Hypothesis and Its Critics}.
                Journal of Economic Perspectives, 17(1), 59-82.
	\bibitem{malkiel2} Malkiel B.~G.~(2005). {\it Reflections on the Efficient Market Hypothesis: 30 Years Later}.
		The Financial Review 40, 1--9.
	\bibitem{sewell} Sewell~M.~(2011). {\it History of the Efficient Market Hypothesis}. UCL Dep.\ of Computer Science Research Note RN/11/04.
	\bibitem{buffet} Buffett W.~(1983), {\it The superinvestors of Graham-and-Doddsville}.
		Hermes: the Columbia Business School Magazine. 4-15.
	\bibitem{grossman} Grossman~S..~and Stiglitz~J.~E.~(1980). {\it On the Impossibility of Infotmationally Efficient Markets}.
		The American Economic Review, 70(3) 393-408.
	\bibitem{box} Box G.~E.~P., Jenkins G.~M., Reinsel G.~C.~and Ljung G.~(2016). {\it Time seires Analysis: Forecasting and Control}, 5th Ed.
		Wiley.
	\bibitem{shadbolt} Shadbolt J.~and Taylor J.~G.~(2002).
		{\it Neural Networks and the Financial Markets: Predicting, Combining and Portfolio Optimisation}. Springer.
	\bibitem{atsalakis} Atsalakis G.~S.~and Valavanis K.~P.~(2009). {\it Surveying stock market forecasting techniques
		- Part II: Soft computing methods}. Exp.\ Sys.\ Appl.\ 36, 5932-5941.
	\bibitem{wang} Wang J.-Z., Wang J.~J., Zhang Z.-G.~and Guo S.-P.~(2011). {\it Forecasting stock indices with back propagation neural network}.
		Exp.\ Syst.\ Appl.\ 38, 14346-14335.
	\bibitem{oliveira} Oliveira F.~A., Nobre C.~N.~and Z\'arate L.~E.~(2013), {\it Applying Artificial Neural Networks to prediction of stock
		price and improvement of the directional prediction index - Case study of PETR4, Petrobras, Brazil}. Exp.\ Syst.\ Appl.\ 40, 7595-7606.
	\bibitem{chong} Chong E., Han C.~and Park F.~C.~(2017). {\it Deep learning networks for stock market analysis and prediction:
		Methodology, data representations and case studies}. Exp.\ Syst.\ Appl.\ 83, 187-205.
	\bibitem{lo3} Lo A.~W.~and MacKinlay A.~C.~(2002). {\it A Non-Random Walk Down Wall Street}. Princeton University Press. pp.~3-11.
	\bibitem{prechter} Prechter Jr, R.~R.\ and Parker, W.~D.~(2007). {\it The financial/economic dichotomy in social behavioral
		dynamics: The socionomic perspective}. The Journal of Behavioral Finance, 8(2), 84–108 .
	\bibitem{peng} Peng C.-K., Buldyrev S.~V., Havtin S., Simons M., Stanley H.~E.~and Goldberger A.~L.~(1994).
		{\it Mosaic organization of DNA nucleotides}. Physical Review E 49(2), 1685-1689.
	\bibitem{podobnik} Podobnik B.~and Stanley H.~E.~(2008). {\it Detrended Cross-Correlation Analysis: A New Method
		for Analyzing Two Nonstationary Time Series}. Physical Reviwe Letters 100, 084102-1--084102-4.
	\bibitem{lima} Lima N.~F., Fernandes L.~H.~S., Jale J.~S.~, Mattos Neto P.~S.~G., Sto\v ci\'c T., Sto\v ci\'c B. and Ferreira T.~A.~E.~(2018).
		{\it Long-term correlations and cross-correlations in IBovespa and constituent companies}. Physica A 492, 1431-1438.
	\bibitem{mantegna2} Mantegna R.~N. and~Stanley H.~E.~(1996). {\it Turbulence and financial markets}. Nature 383, 46-49.
	\bibitem{mantegna1} Mantegna R.~N. and~Stanley H.~E.~(1994). {\it Stochastic Process with Ultraslow Convergence to a Gaussian:
		The Truncated L\'evy Flight}. Physcal Review Letters 28(22), 2946-2949.
	\bibitem{adams} Adams W.~J.~(2000). {\it The Life and Times of the Central Limit Theorem}, 2nd Ed. History of Mathematics 35.
		The American Mathematical Society.
	\bibitem{nelis} McNelis P.~D.~(2005). {\it Neural Networks in Finance}, Springer.
	\bibitem{silva} da Silva I.~N~, Spatti D.~H.~, Flauzino R.~A.~, Liboni L.~H.~B.~and Alves S.~F.~R.~ (2017).
		{\it Artificial Neural Networks}. Springer Int.\ Pub.\ (Switzerland).
	\bibitem{hornik} Hornik K.~(1991). {\it Approximation capabilities of multilayer feedforward networks}. Neural Networks 4(2), 251--257.
	\bibitem{calude} Calude C.~S.\ and Longo G.~(2017). {\it The Deluge of Spurious Correlations in Big Data}. Foundations of Science 22, 595-612.
	\bibitem{rochafilho} Rocha Filho T.~M.~, Oliveira Z.~R~ T.~, Malbouisson L.~A.~C.~, Gargano R.~and Soares Neto J.~J.~ (2003).
		{\it The Use of Neural Networks for Fitting Potential Energy Surfaces: A Comparative Case Study for the H${}_3^+$ Molecule}.
		International Journal of Quantum Chemistry, 95, 281-288.
	\bibitem{numrecip} Press W.~H.~, Teukolsky S.~A.~, Vetterling W.~T.~and Flannery P.~ (1992).
		{\it Numerical Recipes}, 2nd edition. Cambridge University Press.
	\bibitem{adam} Kingma D.~P.~and Ba J.~L.~ (2017). {\it ADAM: A Method for Stochastic Optimization}. arXiv:1412.6980 [cs.LG].
	\bibitem{smith} Smith G.~(2018). {\it The AI delusion}. Oxford Univ.\ Press.
\end{thebibliography}
\end{document}